\documentclass[amsmath,amssymb,aps,superscriptaddress,reprint]{revtex4-1}
\usepackage{graphicx} 
\usepackage{dcolumn} 
\usepackage{bm} 
\usepackage{hyperref} 
\usepackage{longtable} 
\usepackage[T1]{fontenc}
\usepackage{times}
\usepackage{xcolor}
\usepackage[normalem]{ulem}
\usepackage{romannum}

\DeclareMathOperator{\tr}{Tr}

\begin{document}
\title{Predicting colloidal crystals from shapes via inverse design and machine learning}

\author{Yina \surname{Geng}}
\altaffiliation{These authors contributed equally to this work.}
\affiliation{Department of Physics, University of Michigan, Ann Arbor MI 48109,
USA}
\author{Greg \surname{van Anders}}
\altaffiliation{These authors contributed equally to this work.}
\affiliation{Department of Physics, University of Michigan, Ann Arbor MI 48109, USA}
\author{Sharon C. \surname{Glotzer}}
\affiliation{Department of Physics, University of Michigan, Ann Arbor MI 48109, USA}
\affiliation{Department of Chemical Engineering, University of Michigan, Ann
Arbor MI 48109, USA}
\affiliation{Department of Materials Science and Engineering, University of
Michigan, Ann Arbor MI 48109, USA}
\affiliation{Biointerfaces Institute, University of
Michigan, Ann Arbor MI 48109, USA}
\email{grva@umich.edu, sglotzer@umich.edu}

\date{\today}

\begin{abstract}
A fundamental challenge in materials design is linking building block attributes to crystal structure.
  Addressing this challenge is particularly difficult for systems that exhibit
  emergent order, such as entropy-stabilized colloidal crystals. We combine recently developed techniques in inverse design with machine learning to
  construct a model that correctly classifies the crystals of more than ten thousand polyhedral shapes into 13
  different structures with a predictive accuracy of 96\% using only two
  geometric shape measures. With three measures, 98\% accuracy is achieved. We test our model on 
  previously reported colloidal crystal structures for 71 symmetric polyhedra and obtain 92\% accuracy. Our findings (1) demonstrate
  that entropic colloidal crystals are controlled by surprisingly few parameters, 
  (2) provide a quantitative model to predict these crystals solely from the geometry of their building
  blocks, and (3) suggest a prediction paradigm that easily generalizes to other self-assembled materials.
\end{abstract}

\maketitle

A holy grail for materials researchers is the ability to predict crystal
structures solely from knowledge about the constituent atoms, molecules, or
particles, without the need for simulations or experiments. Many researchers
have attempted this quest; a well-known example is the early effort of Pauling
to predict crystal structures from atoms based solely on their atomic radii
\cite{pauling_principles}.  Pauling's rules have since been adopted to program
the assembly of DNA-functionalized nanospheres into colloidal crystals
isostructural to those Pauling considered for atoms, as well as ones with no
atomic counterpart\cite{mirkin1996dna, mirkindnadesign,Gangsuperlattices,
Gangdiamond}. Prediction of crystals from molecules or anisotropic particles is
substantially harder \cite{entint,epp,sangminclathrate}.  In 2012, a study of
145 different polyhedrally shaped particles and their entropy stabilized
crystals provided sufficient data to discover a correlation between coordination
number and isoperimetric quotient (IQ), a measure of the roundness of a particle
\cite{zoopaper}. The study found that knowledge of the coordination number in
the dense fluid (a simple observable in simulations) and the particle IQ allows
one to predict whether that fluid of particles will crystallize and, if so,
whether it will form a liquid crystal mesophase, a medium-coordination crystal,
or a close-packed crystal (including topologically close-packed phases). Despite
being only partially predictive, that was the state of the art in 2012 and in
the half dozen years since, despite continued discoveries of colloidal crystals
that add to the knowledge base. 

Here we leverage two recent important computational advances to enable a much
higher level of predictiveness of colloidal crystal structures from particle
shape. One advance is the inverse design methodology of digital
alchemy\cite{digitalalchemy,engent}, a molecular simulation method in which
particle attributes are treated as thermodynamic variables in a generalized, or extended, 
thermodynamic ensemble \cite{extendedEnsemble, optimalPackingSphere}. 
Digital alchemy simulations produce optimal particle attributes for target thermodynamic phases,
including complex colloidal crystals with very large unit cells.\cite{engent} The other
advance is the application of machine learning methods to materials problems.
\cite{Meredig2014,Ghiringhelli2015,AgrawalAPLM2016,Ferguson2016,
Materiomics2017,Ramprasad2017}
Machine learning can discover hidden correlations in large datasets, providing
clues to the long-sought relationship between building block and structure.  By
combining these two approaches to thermodynamic systems of hard, polyhedrally
shaped particles, we find an empirical but highly predictive model for
entropically ordered colloidal crystals from solely geometric measures of their
constituent particles. Our model is capable of predicting 13 different
entropically stable crystal structures formed by millions of different colloidal
polyhedra with 96\% fidelity using just two geometric measures of particle shape
(see Fig.\ 1 for an illustration of our approach). With three measures, 98\%
fidelity is achieved. Though far from a first principles theory, this model can
be used immediately to inform experiments and select building blocks for
self-assembling nanoparticle superlattices and colloidal crystals.

To construct the predictive model we performed Alchemical Monte Carlo (Alch-MC)
simulations based on the Digital Alchemy framework \cite{digitalalchemy}, using
an implementation \cite{engent} that extends the Hard Particle Monte Carlo
(HPMC) plugin \cite{hpmcplug} for HOOMD-Blue \cite{hoomdblue} to generalized
thermodynamic ensembles that include particle shape change. We simulated
$NVT\mu$ ensembles at constant temperature $T$, fixed volume $V$, and chemical
potential $\mu=0$ for 13 target structures reported previously in entropic
self-assembly: simple cubic (SC), body-centered cubic (BCC), face-centered cubic
(FCC), simple chiral cubic (SCC), hexagonal (HEX-1-0.6), diamond (D), graphite
(G), honeycomb (H), body-centered tetragonal (BCT-1-1-2.4), high-pressure
Lithium (Li), $\beta$-Manganese ($\beta$-Mn), $\beta$-Uranium ($\beta$-U), and
$\beta$-Tungsten ($\beta$-W). The variable $\mu$ is conjugate to the shape
variable that is allowed to fluctuate in the simulation. We placed a minimum of
$N = 100$ particles in a periodic simulation box, with the exact number chosen
to be a multiple of the number of particles in the unit cell of one of the 13
target structures.  Particle shapes were initialized with as many as 64 vertices
randomly generated to create a convex shape. Monte Carlo (MC) sweeps were
performed to allow particle translations, rotations, and shape moves via vertex
re-location. For each shape move, we (i) moved a vertex, (ii) resized the trial
shape to unit volume, (iii) checked if the move induced any particle overlaps,
and then (iv) accepted the move based on the Boltzmann factor as described in
Ref.\ \cite{digitalalchemy}. Translation and rotation moves followed standard
procedures (see, e.g., Refs.\
\cite{amirnature,escobedo,dijkstratcube,trunctet,zoopaper,entint,epp}).  

We slowly compressed the target crystal structure comprised of a randomly
generated shape to the target packing fraction, with springs of spring constant
$1000$ (where energy is specified in units of $k_\mathrm{B}T$ and length units
are set by the particle size) at each node to maintain the integrity of the
structure during the compression phase. In this initial stage, it is highly
unlikely that the structure is thermodynamically stable, so fixing the positions
of the centers of the particles over an initial set of MC steps allows the
system to explore shape space during compression without falling apart. After
reaching the target packing fraction, we logarithmically relaxed the spring
constant to zero.  We then further evolved the system at fixed packing fraction
$\eta$ for $1\times10^6$ (BCC-$\eta=0.6$, FCC-$\eta=0.6$, SC-$\eta=0.6$,
diamond-$\eta=0.6$, honeycomb-$\eta=0.65$, graphite-$\eta=0.65$,
HEX-1-0.6-$\eta=0.7$, BCT-1-1-2.4-$\eta=0.7$,  Li-$\eta=0.65$) or $2\times10^6$
(SCC-$\eta=0.7$) or  $8\times10^6$ ($\beta$-W-$\eta=0.6$) or
$3.6\times10^7$($\beta$-Mn-$\eta=0.6$) or $1\times10^8$ ($\beta$-U-$\eta=0.6$)
MC sweeps. For each target crystal structure, we performed 10 independent
simulations and analyzed the shapes in the final $5\times10^5$ sweeps. The
output of the Alch-MC simulation procedure for each of the 13 crystal structures
is a family of optimal shapes (that is, shapes that minimize the system free
energy) with shape measures that fluctuate about some average value.

We calculated 10 measures of shape motivated by prior works studying the
relationship between particle geometry and self-assembly
behavior.\cite{zoopaper} The shape measures calculated for each polyhedral shape
are: the cosine of the average dihedral angle $\theta_d$; the number of facets
$N_f$; the determinant of the moment of inertia $I$; the trace of the moment of
inertia; the ratio of the circumsphere radius to the insphere radius; IQ; the
asphericity $\alpha$; and the real and imaginary parts of the chiral parameter
$\chi$.  To calculate $\cos \theta_d$, facets with area $a_f>a_f^*$ (we use
$a_f^*=0.02$ but our results are not sensitive to changes in $a_f^*$) were
clustered by their normal vector using the DBSCAN \cite{dbscan} scikit-learn
module \cite{scikit-learn}.  Clustered facets are represented by area-weighted
average normals.  We computed the clustered-facets-area-weighted cosine of the
angle between average normals of neighboring clustered facets in a polyhedron.
The number of facets is the number of clustered facets using DBSCAN.  The
isoperimetric quotient $IQ = 36\pi v^2/s^3$, where $v$ is the volume and
$s$ is the surface area of a polyhedron).  The
asphericity $\alpha = Rs/3v$, where $R$ is the integrated mean curvature
normalized by $4\pi$. Finally, $\chi$ is the lowest order measure of the degree
of chirality of a molecule \cite{harris1999molecular} and it is zero for achiral
molecules. $\chi \propto \sum\limits_{mn}
C(234;mn)\rho_{2m}\rho_{3n}\rho_{4,m+n}^*$, where $C(234;mn)$ are the
appropriate Clebsch-Gordan coefficients and $\rho_{lm}$ are mass-weighted
distance moments with $\rho_{lm} = \sum\limits_{\tau\in T}
|\bold{r}_{\tau}|^lY_{lm}(\theta_{\tau}, \phi_{\tau})$, where the sum is over
atoms $\tau$ in the molecule $T$. We calculated $\chi$ for a polyhedron by
assuming a unit mass atom at each vertex of its vertices.

Altogether, this procedure produced the dataset to be interrogated using machine
learning (ML), with the aim of identifying salient geometric criteria for
predicting structure.  We utilized the random forest ML classification technique
in scikit-learn.  Random forest builds a large collection of decorrelated trees
and then averages them to improve accuracy and control over-fitting.  The
technique is capable of high predictive accuracy and is applicable in
high-dimensional problems such as this one with highly correlated variables.
Here the 10 shape measures are provided as input and the 13 crystal structures
as produced as output.

Table 1 reports correlations found for the $\beta$-Mn structure between the 10
tested  geometric measures applied to 1000 free-energy minimizing shapes
randomly selected from among the millions of shapes generated by the Alch-MC
simulations. Tables reporting correlations for all other structures studied can
be found in the Supplementary Materials, as Tables S1-S12. The ML analysis (see
Fig.\ 2a) of low density data indicates that two shape features -- ($\tr(I)$)
and ($\langle\cos(\theta_d)\rangle$) --- are the strongest predictors of the
shape--structure relationship. Fig.\ 2c shows shape distributions for the 13
candidate structures using $\tr(I)$ and $\cos(\theta_d)$.

The
self-assembly of shapes that are now understood to be far from optimal
\cite{engent} has been reported in the literature.\cite{escobedo,zoopaper}.
To develop a model that also correctly predicts assemblies for highly
suboptimal shapes, we run additional simulations at other densities to broaden
our data set. We further generate shape distributions at high densities with different number of shape vertices (FCC: number of vertices=32, $\eta$=0.65; 32, 0.7; 32, 0.8; 32, 0.9; 32, 0.95; 32, 0.99; 10, 0.6; 12, 0.6; 14, 0.6; 20, 0.6;
50, 0.6; 120, 0.6; BCC: 6, 0.6; 14, 0.64), shape distributions are shown in Fig.\ S1. From this we find that distributions
of particle shapes are clustered by geometry, and several structures exhibit
multiple free energy basins. The existence of these basins indicates that for
those structures multiple distinct particle shapes provide good candidate shapes
for self-assembly.

We further note that shape distributions vary in form by
structure. For example, G and H structures exhibit a narrow range of
$\cos(\theta_d)$ relative to the distributions for SC, SCC, BCC, BCT, and FCC.
Conversely SC, SCC, BCC, BCT, and FCC all exhibit a narrow range of $\tr(I)$
relative to the distributions for G and H. We expect that this
  relative sensitivity is an important consideration in the synthesis of
appropriately shaped particles for self-assembly. Moreover,
Fig.\ 2b shows regions that BCC, FCC,
$\beta$-W, $\beta$-U, and $\beta$-Mn distributions locate tightly together in
shape space with some overlap. The existence of this overlap accords with prior
work \cite{zoopaper} that found some hard polyhedra spontaneously self-assemble
more than one structure (e.g.\ $\beta$-Mn and $\beta$-U).

Fig.\ 2d shows the two-parameter empirical model, obtained from the Alch-MC
simulation data, that maps particle shape to structure. We also extracted
thermodynamically optimal shapes from the peaks of the shape distributions in
the two salient parameters. Fig.\ 3 shows peak positions, and inset particle
images show representative particle shapes. Remarkably, the two-parameter model
predicts structure from shape with 96\% accuracy. Inclusion of $\alpha$
increases the accuracy of structure prediction to 98\%.

As a further test of the model's predictive ability, we compared our model's
predictions against the previously reported self-assembly behavior
\cite{zoopaper} of 81 known crystal-forming polyhedra, in which 71 shapes formed
target structures included in our set of 13.  Contained in that
  set are the cube \cite{henzieSuperlattices, rossiCubicCrystal}, truncated cube
  \cite{MirkinSuperlattices, henzieSuperlattices}, octahedron
  \cite{MirkinSuperlattices, henzieSuperlattices}, truncated octahedron
  \cite{henzieSuperlattices}, cuboctahedron \cite{henzieSuperlattices}, and
  rhombic dodecahedron \cite{MirkinSuperlattices}, each of whose self-assembled
colloidal crystal structures have been shown in experiments to match the
simulated structures.  In Ref. ~\cite{zoopaper}, shapes were self-assembled
into crystals without the use of alchemical variables, and thus the shapes are
not necessarily the optimal ones for the obtained crystal structure. For that
reason, we expect the accuracy of our predictions to be lower than that obtained
for optimal shapes.  We found that the two-parameter model correctly predicts
the crystal structure formed by 65 shapes in the 2012 study with a fidelity of
91.55\%. The model is, of course, unable to correctly predict the structure for
shapes that formed structures not included in our set of 13, and also failed to
predict structures where multiple structures were reported. Model predictions
and shapes are given in Fig.\ 4. 

Our coupling of of inverse design and machine learning techniques to create a
purely geometric, two-parameter, empirical model that predicts the self assembly
of colloidal crystals of convex polyhedra with a fidelity of 96\% is a
considerable advance over the two-parameter, empirical model presented in Ref.\
\cite{zoopaper}.  Nevertheless, there is room for further improvement. First,
our model was based on considering only 10 different shape features. It is
possible that there are other shape features that could prove better predictors
of self-assembly behavior. To facilitate investigations into this possibility,
we have made raw shape data available online at
\url{https://deepblue.lib.umich.edu/data/concern/generic_works/6q182k84r?locale=en}.
We encourage the community to use the data to look for increasingly accurate
predictive models.  Second, in regions of shape space that are either sparsely
populated by our data and so yield poor statistics, or in which different
structures are densely clustered, the model may fail to predict the correct
structure. We encourage the community to add data to the online data set for
additional shapes and structures beyond those considered here.

The fact that as few as two geometric criteria are sufficient to predict
structure formation in crystals with as many as 30 ($\beta$-U) particles in a
unit cell suggests that appropriate geometric criteria might be useful for
predicting order in colloidal systems with other forces at play, including
hydrogen bonding between DNA-programmable colloidal shapes, van der Waals
attraction between ligands or between particle cores, depletion interactions,
and so on. Alch-MC or other inverse design methods can treat any of these cases
to produce optimized forces as well as shapes. Nevertheless, our predictions
should extend to experimental systems of anisotropic colloids and nanoparticles
in which entropy plays a major role.\cite{colloidalmatter,entint}

\begin{figure*}
  \includegraphics[width=\textwidth]{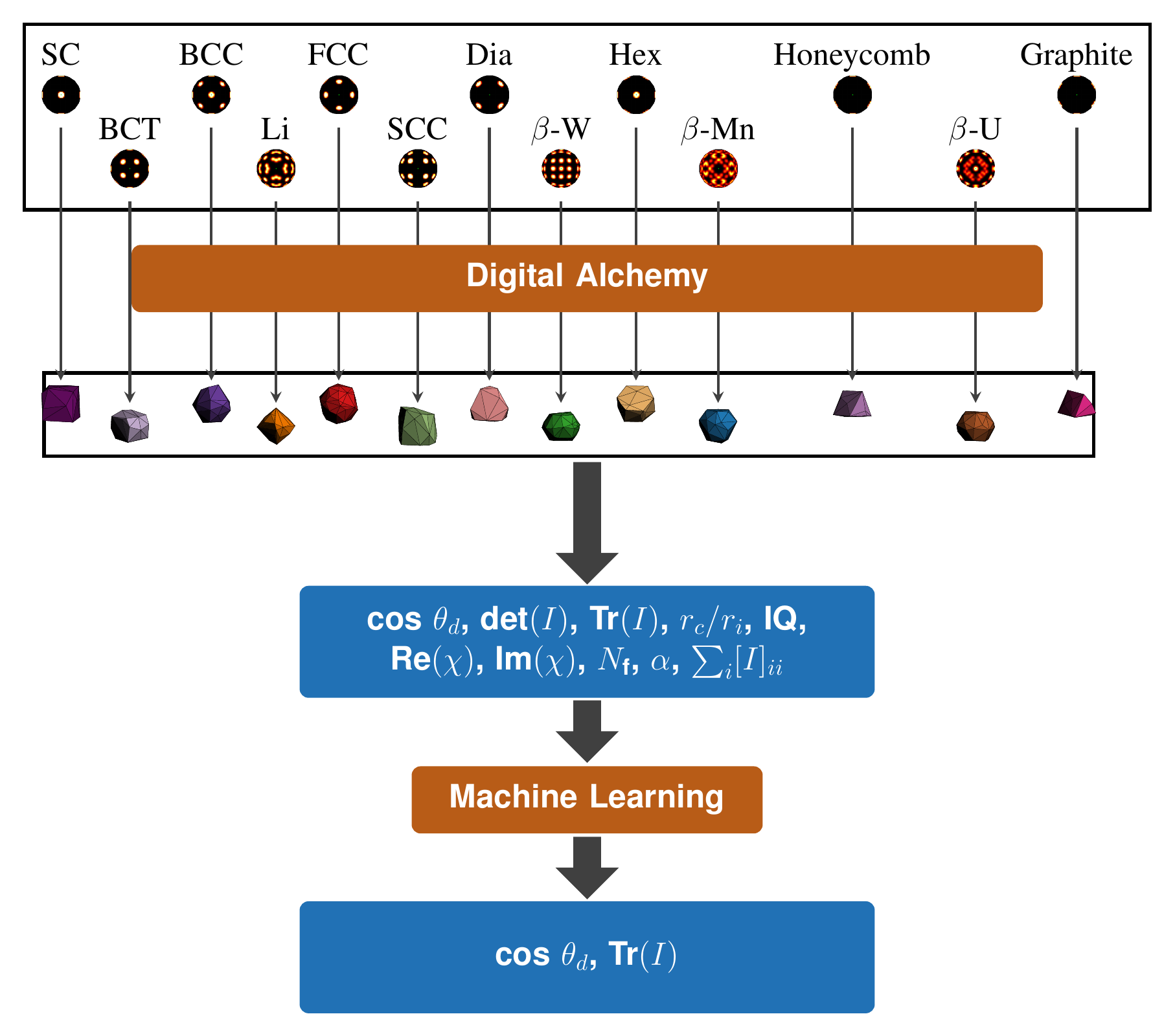}
  \caption{
    We use the Digital Alchemy inverse materials design approach to find optimal
    and near-optimal hard, convex, colloidal, polyhedral shapes for 13 target
    structures. We use the Random Forest technique from machine learning to
    classify shapes. We find that, of 10 measures of shape, two -- the dihedral
    angle ($\cos(\theta_d)$) and the trace of the moment of inertia tensor
    ($\tr(I)$) -- are sufficient to predict the self assembly behavior of a
    shape.
  }
\end{figure*}

\begin{figure*}
  \includegraphics[width=\textwidth]{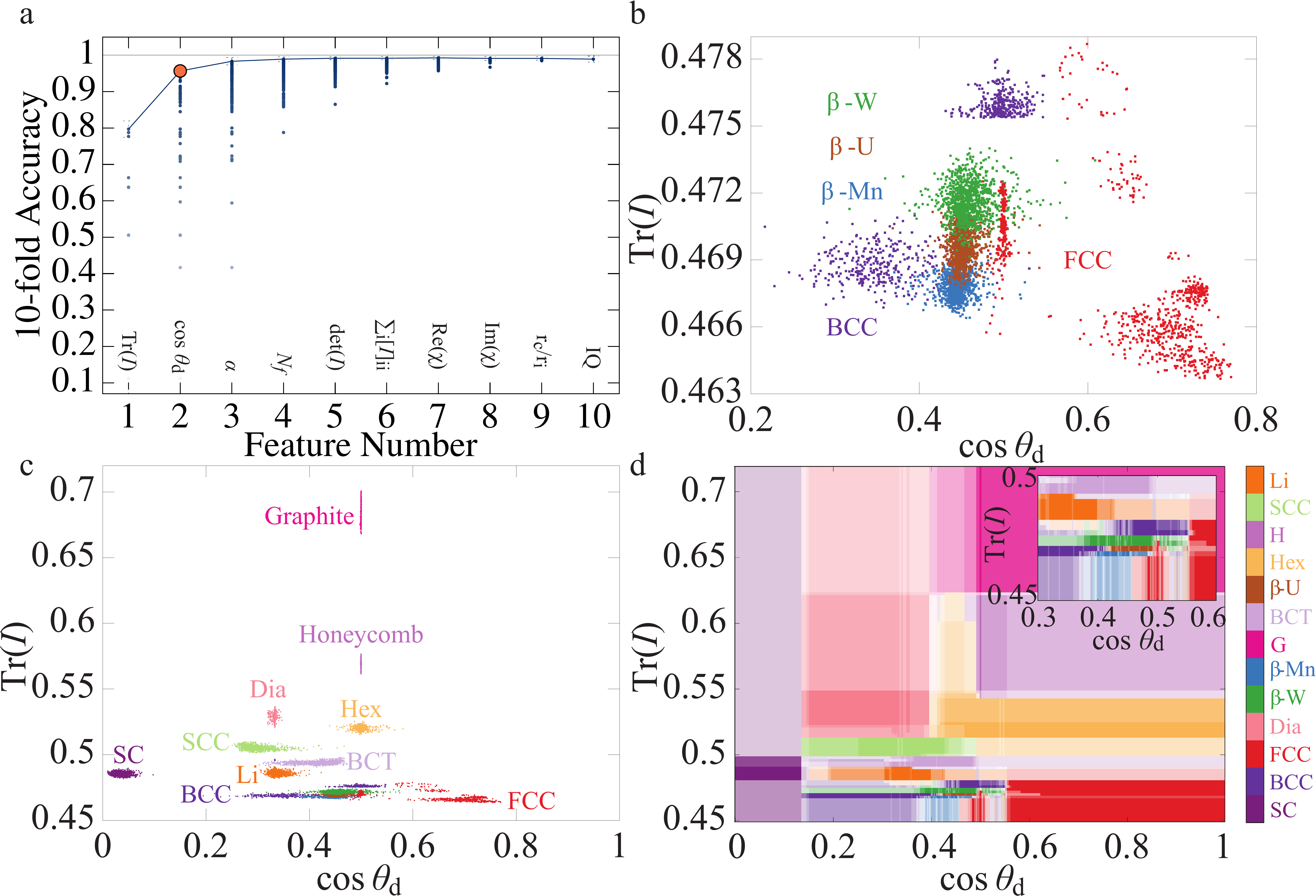}
  \caption{
    (a) From low density shape distributions produced via Alch-MC, we classify shapes using
    combinations of 10 geometric criteria via the random forest method from
    machine learning. We find that two shape features ($\cos(\theta_d)$ and 
    $\tr(I)$) give greater than 91\% accuracy in predicting structure.
    (c) Shape distributions from Alch-MC plotted as a function of the two
    primary shape features (Including both low density and high density shapes). Each mark represents an observed shape, and is
    colored by its corresponding crystal structure. Inset shows regions of shape
    space in which different structures are densely clustered.
    (b) Zoom in of the densely distributed structure region in (c).
    (d) Model prediction probability based on particle geometry of self-assembly behavior of
    hard convex polyhedra that crystallize into 13 target structures.
  }
\end{figure*}
\begin{figure*}
  \includegraphics[width=\textwidth]{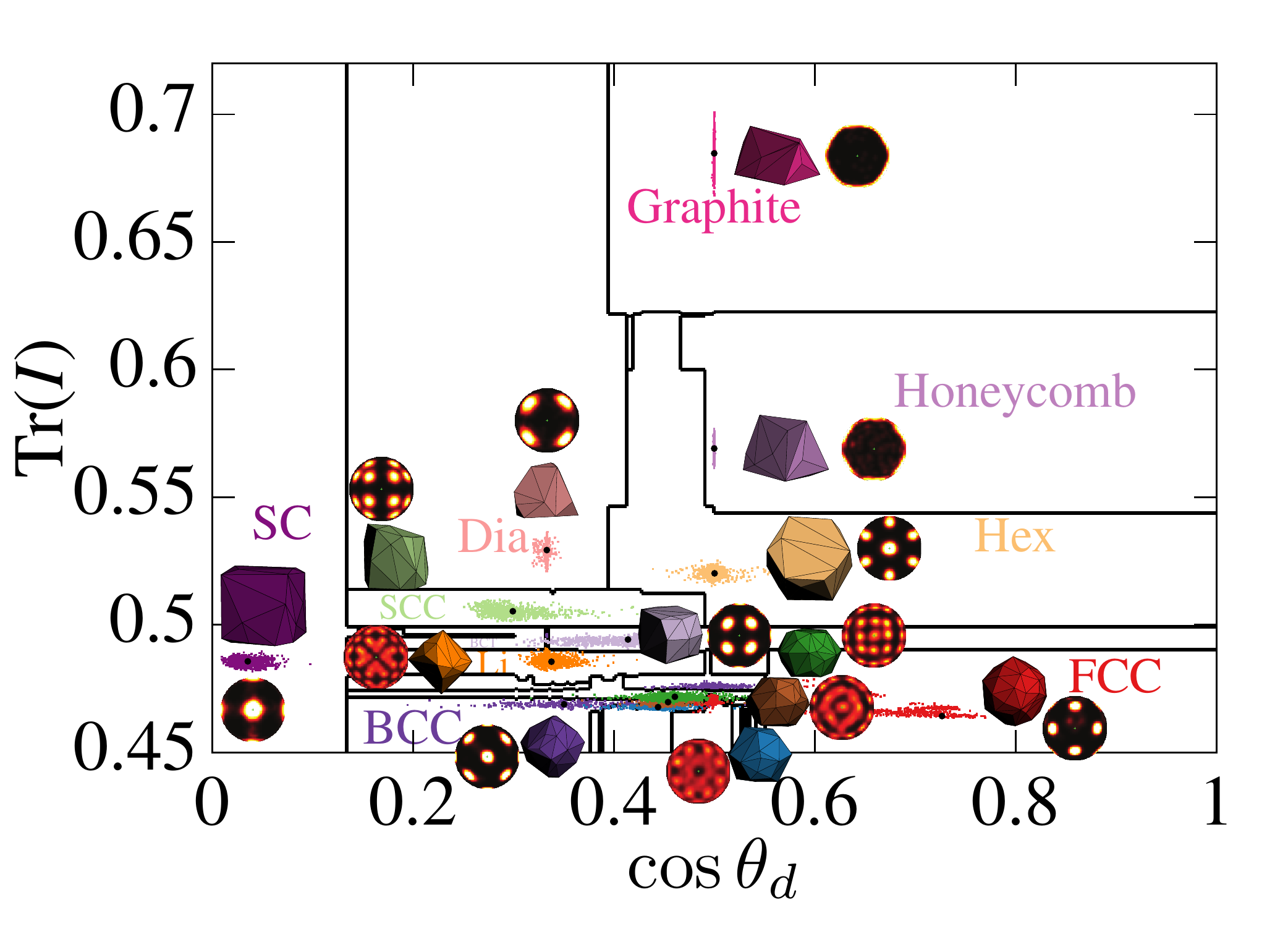}
  \caption{
    Model prediction of self-assembly of hard, convex, colloidal polyhedra.
    Marks indicate observed particle shapes and are colored by structure. Black
    marks indicate shape distribution peaks corresponding to thermodynamically
    optimal particle shapes. Example near-optimal particle shapes are shown.
  }
\end{figure*}
\begin{figure*}
  \includegraphics[width=\textwidth]{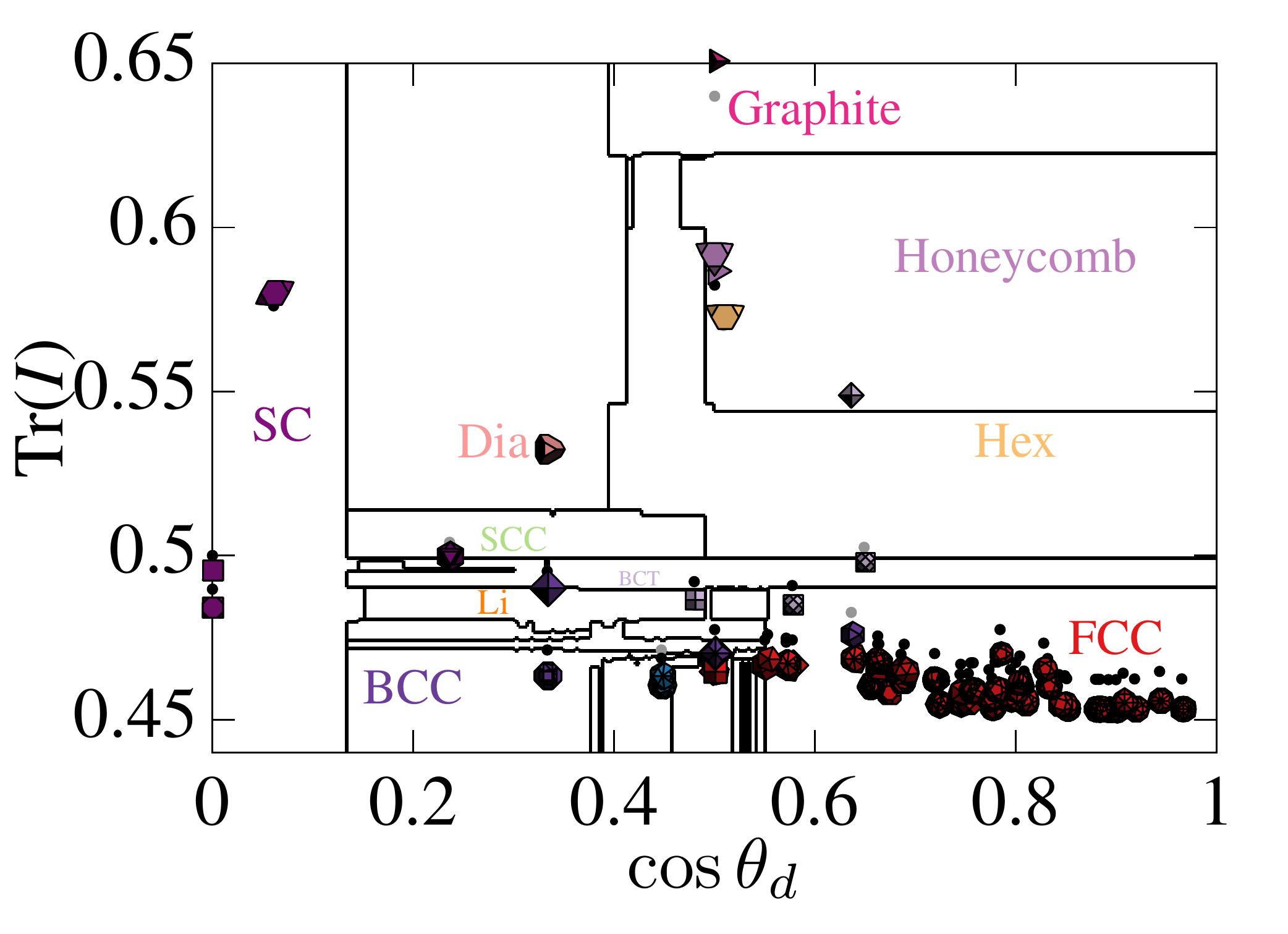}
  \caption{
    Test of empirical model prediction against previously reported dataset
    describing crystal self-assembly of 71 convex polyhedra. Model correctly
    predicts precise crystal structure observed in self-assembly in 65 of 71
    cases (91.55\%).
  }
\end{figure*}

\begin{table*}
  \begin{center}
    \begin{tabular}{ |p{1.75cm}|p{1.25cm}|p{1.25cm}|p{1.25cm}|p{1.25cm}|p{1.25cm}|p{1.25cm}|p{1.25cm}|p{1.25cm}|p{1.25cm}|p{1.25cm}|}
      \hline
      Correlation & $\cos \theta_d$ & $r_c/r_i$ & det($I$) & IQ & Re($\chi$) & Im($\chi$)   & $N_f$ & $\alpha$ & $\sum_i[I]_{ii}$ & Tr($I$) \\
      \hline
     $\cos \theta_d$   &   1  & 0.03  &  0.02 & -0.1 & 0.01 & -0.09 & 0 & 0.08 & 0.02 & 0.02  \\
      \hline
     $r_c/r_i$   &  0.03 &  1 &  0.51  & -0.23 &  0.04 &  -0.06 &  -0.06 &  0.3 &  0.52 &  0.54 \\
      \hline
    det($I$)   &  0.02   & 0.51 &  1 & -0.51 &  0.09 &  -0.03 &  -0.07 &  0.58 &  1  &  0.99 \\
      \hline
    IQ     &  -0.1   &  -0.23 &  -0.51 &  1 &  -0.02 &  0.27 &  -0.31 &  -0.97 &  -0.5  & -0.5  \\
      \hline
     Re($\chi$)  &  0.01  &  0.04 &  0.09 &  -0.02 &  1 & -0.03 &  0 &  0.02 &  0.08 &  0.08  \\
      \hline
     Im($\chi$)  &-0.09   &  -0.06 &  -0.03 &  0.27 &  -0.03 &  1 &  -0.14&  -0.22 &  -0.03 &  -0.03  \\
        \hline
     $N_f$  &  0&  -0.06 &  -0.07 &  -0.31 &  0 &  -0.14 &  1 &  0.21 &  -0.06 &  -0.06  \\
        \hline
      $\alpha$  &  0.08  &  0.3 &  0.58 &  -0.97 &  0.02 &  -0.22 &  0.21 & 1 & 0.58 &  0.57 \\
        \hline
      $\sum_i[I]_{ii}$  &  0.02 &  0.52 &  1 & -0.5 &  0.08 &  -0.03 &  -0.06 &  0.58 & 1 & 1\\
        \hline
     Tr($I$)  &  0.02 & 0.54 & 0.99 & -0.5 & 0.08 & -0.03 & -0.06 & 0.57 & 1 & 1\\
      \hline
     
    \end{tabular}
  \end{center}
  \caption{Correlation matrix of 10 geometric measures for the $\beta$-Mn structure.}
\end{table*}

\end{document}